\documentclass[journal]{IEEEtran}

\usepackage{cite}

\ifCLASSINFOpdf
  \usepackage[pdftex]{graphicx}
\else
\fi

\usepackage{amsmath,amsfonts,amssymb}
\usepackage{times}
\usepackage{graphicx}
\usepackage{url}
\usepackage{booktabs}
\usepackage{bbm}
\usepackage{diagbox}

\usepackage{array}

\usepackage{multicol}
\usepackage{multirow}

\newcommand{\tabincell}[2]{\begin{tabular}{@{}#1@{}}#2\end{tabular}}

\usepackage{url}

\begin{document}
%
\title{Classification of Large-Scale High-Resolution SAR Images with Deep Transfer Learning}
%
%
%
\author{Zhongling~Huang,
        Corneliu~Octavian~Dumitru,
        Zongxu~Pan,~\IEEEmembership{Member,~IEEE}~
        Bin~Lei,
        and~Mihai~Datcu,~\IEEEmembership{Fellow,~IEEE}~ 
\thanks{This work was supported by the University of Chinese Academy of Sciences (UCAS) Joint PhD Training Program scholarship. \textit{(Corresponding author: Mihai Datcu.)}} 
\thanks{Z. Huang, Z. Pan and B. Lei are with the Aerospace Information Research Institute, Chinese Academy of Sciences, Beijing 100094, China, School of Electronic, Electrical and Communication Engineering, University of Chinese Academy of Sciences, Huairou District, Beijing 101408, China, and also with Key Laboratory of Technology in Geo-spatial Information Processing and Application System, Chinese Academy of Sciences, Beijing, China. (e-mail: huangzhongling15@mails.ucas.ac.cn).} 
\thanks{Z. Huang, C.O. Dumitru and M. Datcu are with the EO Data Science Department, German Aerospace Center (DLR), 82234 Wessling, Germany (e-mail: Corneliu.Dumitru@dlr.de; Mihai.Datcu@dlr.de).} 
}
\maketitle

\begin{abstract}
The classification of large-scale high-resolution SAR land cover images acquired by satellites is a challenging task, facing several difficulties such as semantic annotation with expertise, changing data characteristics due to varying imaging parameters or regional target area differences, and complex scattering mechanisms being different from optical imaging. Given a large-scale SAR land cover dataset collected from TerraSAR-X images with a hierarchical three-level annotation of 150 categories and comprising more than 100,000 patches, three main challenges in automatically interpreting SAR images of highly imbalanced classes, geographic diversity, and label noise are addressed. In this letter, a deep transfer learning method is proposed based on a similarly annotated optical land cover dataset (NWPU-RESISC45). Besides, a top-2 smooth loss function with cost-sensitive parameters was introduced to tackle the label noise and imbalanced classes' problems. The proposed method shows high efficiency in transferring information from a similarly annotated remote sensing dataset, a robust performance on highly imbalanced classes, and is alleviating the over-fitting problem caused by label noise. What's more, the learned deep model has a good generalization for other SAR-specific tasks, such as MSTAR target recognition with a state-of-the-art classification accuracy of 99.46\%.
\end{abstract}


%
\IEEEpeerreviewmaketitle

\section{Introduction}
%
%
%
%
\IEEEPARstart{S}{ynthetic} Aperture Radar (SAR) land use and land cover (LULC) classification is an important step in SAR image interpretation. Recognizing different land covers in SAR images requires an expertise in interpreting the complex scattering characteristics under various circumstances such as different wavelengths, incidence angles and target areas which makes the annotation of large-scale SAR datasets time consuming. With the development of deep learning algorithms, hierarchical features can be extracted automatically. Unsupervised learning approaches, such as deep belief networks (DBNs) \cite{lv2015urban}, and auto-encoders (AEs) \cite{7518668,hzl}, are widely applied to SAR images for classification. With more detailed texture information, high-resolution (HR) PolSAR images are used to learn spatial and polarimetric features jointly via deep convolutional neural networks (DCNNs) for supervised classification which achieves state-of-the-art performance \cite{HRpolsar}.

However, many SAR LULC classification algorithms mainly focus on specific tasks which may not generalize well on other SAR tasks, for example, classifying vegetation and agricultural land covers in farmland areas \cite{rs8080684}, or training an urban-specific network with two scene images in urban areas \cite{8127696}. Although a big volume of SAR data all over the world is acquired every day, it is challenging to automatically interpret land covers in SAR images with such diversity and to obtain a well-generalized model for SAR image understanding.

In this letter, a global dataset of multi-level annotated HR SAR images \cite{dumitru2016land} has been experimented. These SAR images vary in geographic locations, incidence angles, radiometric calibration factors, and ascending or descending satellite orbit branches. With the difficulties of highly imbalanced categories, geographic diversity and noisy annotations, our contributions are as follows:

(1) A very deep residual network with transitive transfer learning method from natural images to remote sensing and then to SAR data is used to understand a large amount of land cover SAR images from all over the world.

(2) A cost-sensitive top-2 smooth combined loss function with cross-entropy loss is proposed to alleviate the imbalanced problem and model bias to mislabeled training samples.

(3) The proposed network is proved to have a good generalization on other SAR datasets and tasks.

In the following, we will firstly introduce the TerraSAR-X (TSX) annotated land cover dataset in Section \ref{2}. Our transfer learning based method, as well as the experimental results will be presented in Section \ref{3} and Section \ref{4}, followed by a conclusion in Section \ref{5}.

\begin{figure*}[!t]
\centering
\includegraphics[width=17cm]{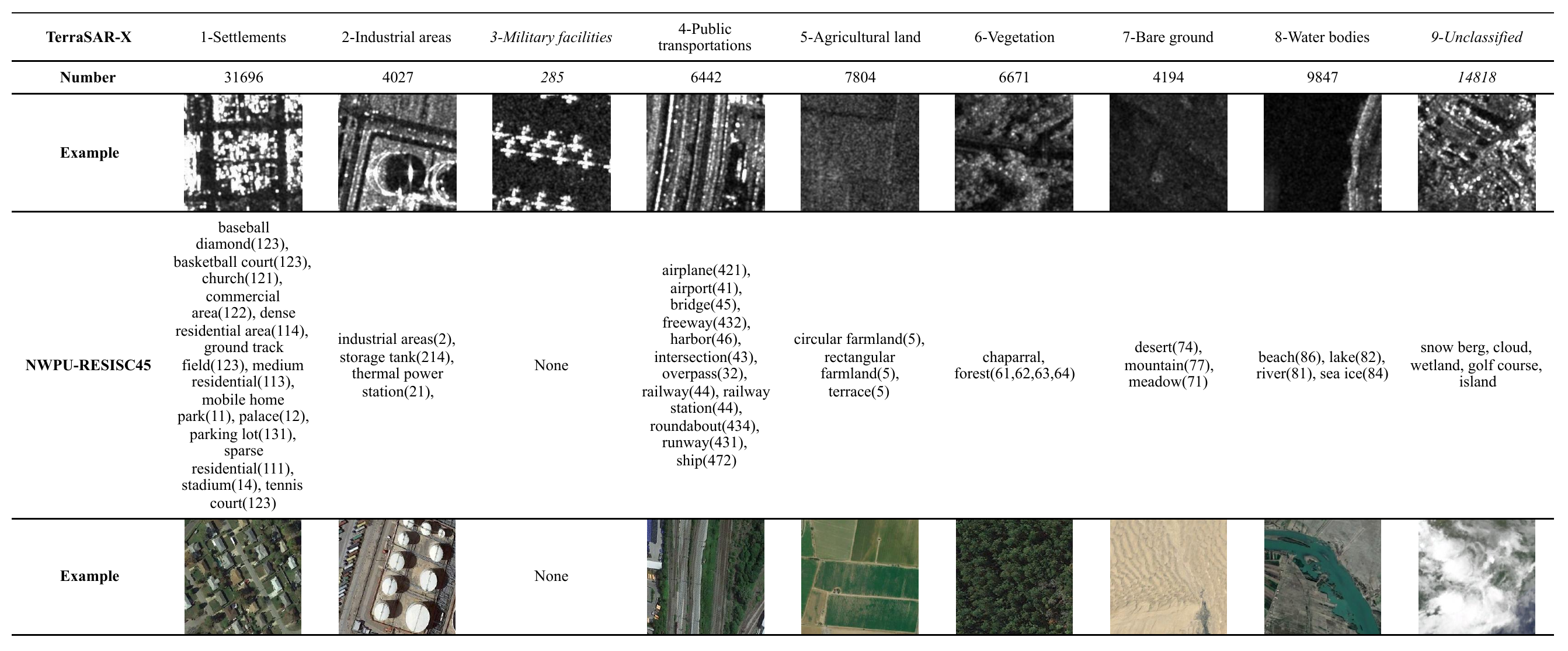}
\caption{Nine kinds of land cover of our annotated level-1 TSX dataset with imbalanced classes. Statistically, 40 categories in NWPU-RESISC45 are semantically similar to the TSX dataset. The number in brackets indicates the assigned label of TSX, which can be indexed in TABLE A.I., Appendix \cite{dumitru2016land}. Due to the small number of patches in "Military facilities" and indistinct label "Unclassified", only seven classes are applied in our experiments.}
\label{Fig1}
\end{figure*}

\section{Datasets}
\label{2}
The HR SAR annotated land cover dataset \cite{dumitru2016land} exploited in this paper has been collected from the X-band TSX instrument \cite{tsxurl}; we used horizontally polarized (HH), multi-look ground range detected (MGD) products which were radiometrically enhanced. The selected SAR images were taken in High Resolution Spotlight mode with a resolution of 2.9 m, acquired with an incidence angle between 20$^{\circ}$ and 50$^{\circ}$, in both descending and ascending pass directions. 

This annotated semantic dataset covers urban and non-urban areas from all over the world, grouped into 46 collections in total. With a pixel spacing of 1.25 m, each SAR image was tiled into non-overlapping patches with a size of 160$\times$160 pixels, resulting in 200$\times$200 m$^2$ on the ground so that a patch covered a typical object on the ground. A single label was assigned to each patch based on the dominating class. With three-level hierarchical annotation, only level-1 labels are used in this paper.

Automatically interpreting the worldwide large-scale SAR dataset is challenging due to the following reasons:

\subsubsection{Highly imbalanced classes}

As shown in Fig. \ref{Fig1}, the imbalanced classes result in a big difference of data volume for each category in level-1 annotation case, mainly due to the fact that the dataset focuses on urban areas. This imbalance problem appears more serious in level-2 and level-3 annotation.

\subsubsection{Geographic diversity}

\begin{figure}[!t]
\center
\includegraphics[width=8cm]{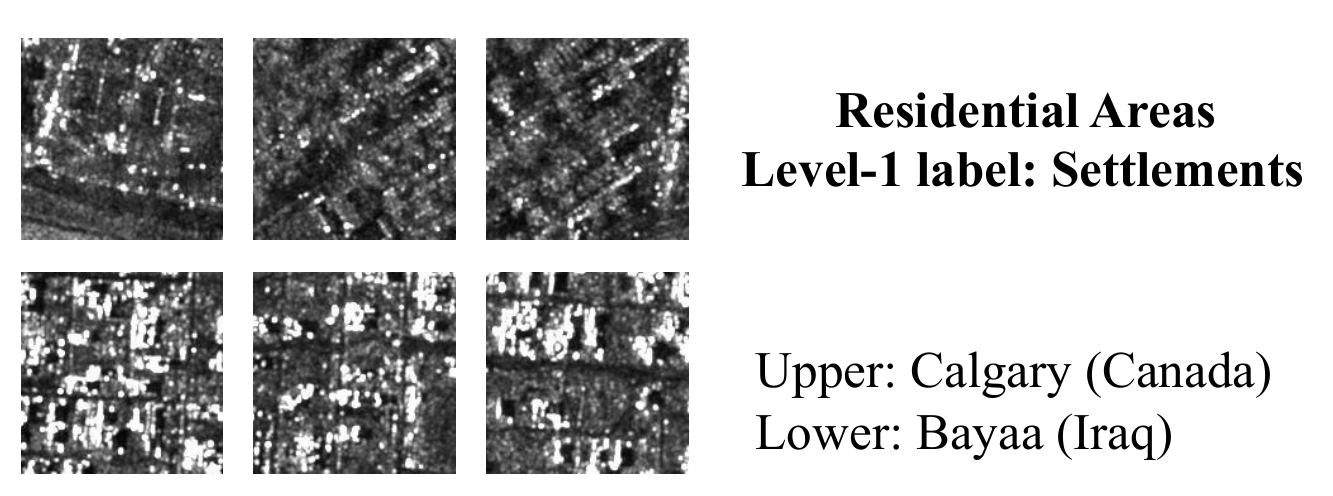}
\caption{The high intra-class diversity of \textit{"low-density residential areas"} in two different cities.}
\label{Fig3}
\end{figure}

Collected world-widely, different architectural styles of cities bring out a large geographic diversity of images. The typical regional characteristics result in high intra-class diversity and inter-class similarity as shown in Fig. \ref{Fig3}. Besides, the acquisition period of these images ranges from 2007 to 2012. During this period of time, some non-constant parameters such as the radiometric calibration factor, will result in different characteristics of the SAR images. 

\begin{figure}[!t]
\includegraphics[width=9cm]{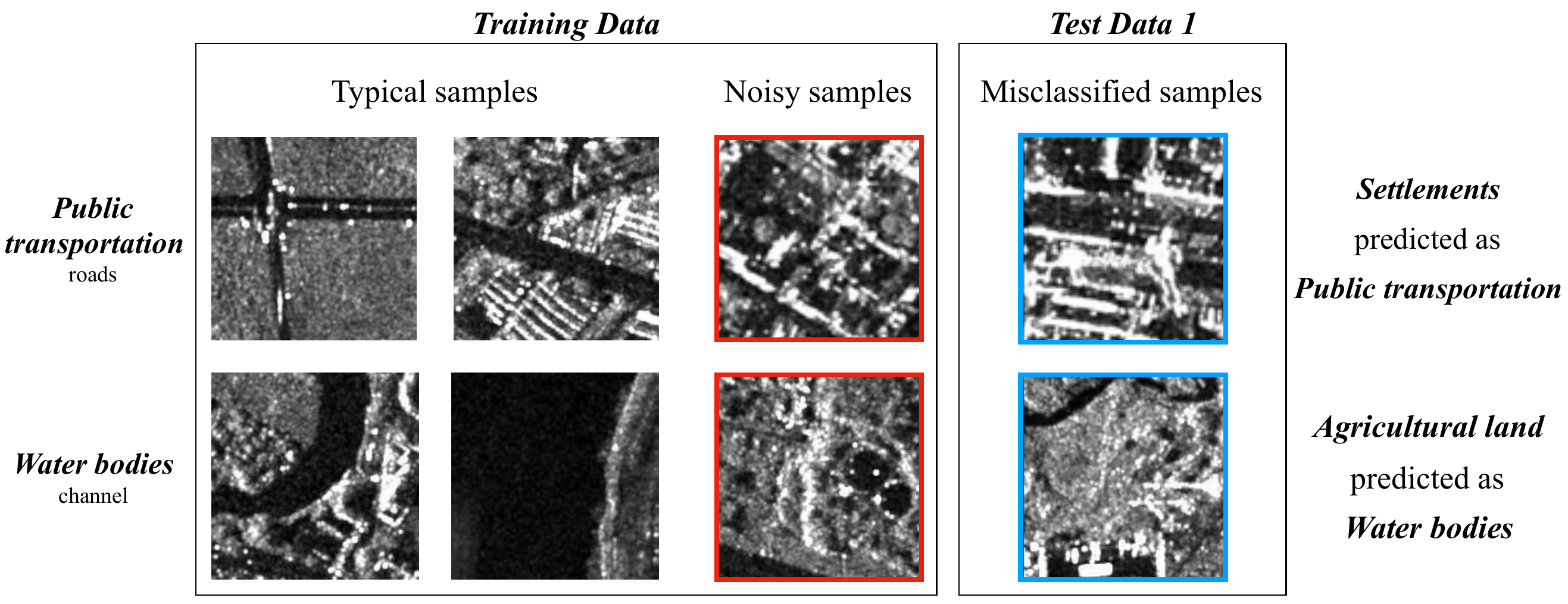}
\caption{Examples of label noise and model bias due to the noisy labels. 
}
\label{Fig6}
\end{figure}

\subsubsection{Label noise}

By covering an area of 200$\times$200 m$^2$ on the ground, sometimes a patch contains more than one land cover type. Even though the annotation was based on the dominating class, there still existed some incomplete annotations or some semantic biases \cite{7926346} which may lead the trained model to over-fitting when using these misleading samples. Fig. \ref{Fig6} gives an example of label noise in our dataset.


Due to the low number of only 285 patches from two collections of \textit{"Military facilities"} and the unlabeled patches of \textit{"Unclassified"}, we only used the remaining seven land cover classes in this paper.

Another remote sensing LULC dataset used in this paper is NWPU-RESISC45 \cite{NWPU}, which has a land cover annotation similar to the seven classes of TSX dataset used in our experiment, shown in Fig. \ref{Fig1}, containing 31,500 images covering 45 scene classes.

\section{Proposed Method}
\label{3}
Facing the difficulties mentioned above, we propose a deep transfer learning based method in this part to tackle the SAR land cover classification issue. 

\begin{figure}[!t]
\centering
\includegraphics[width=8.5cm]{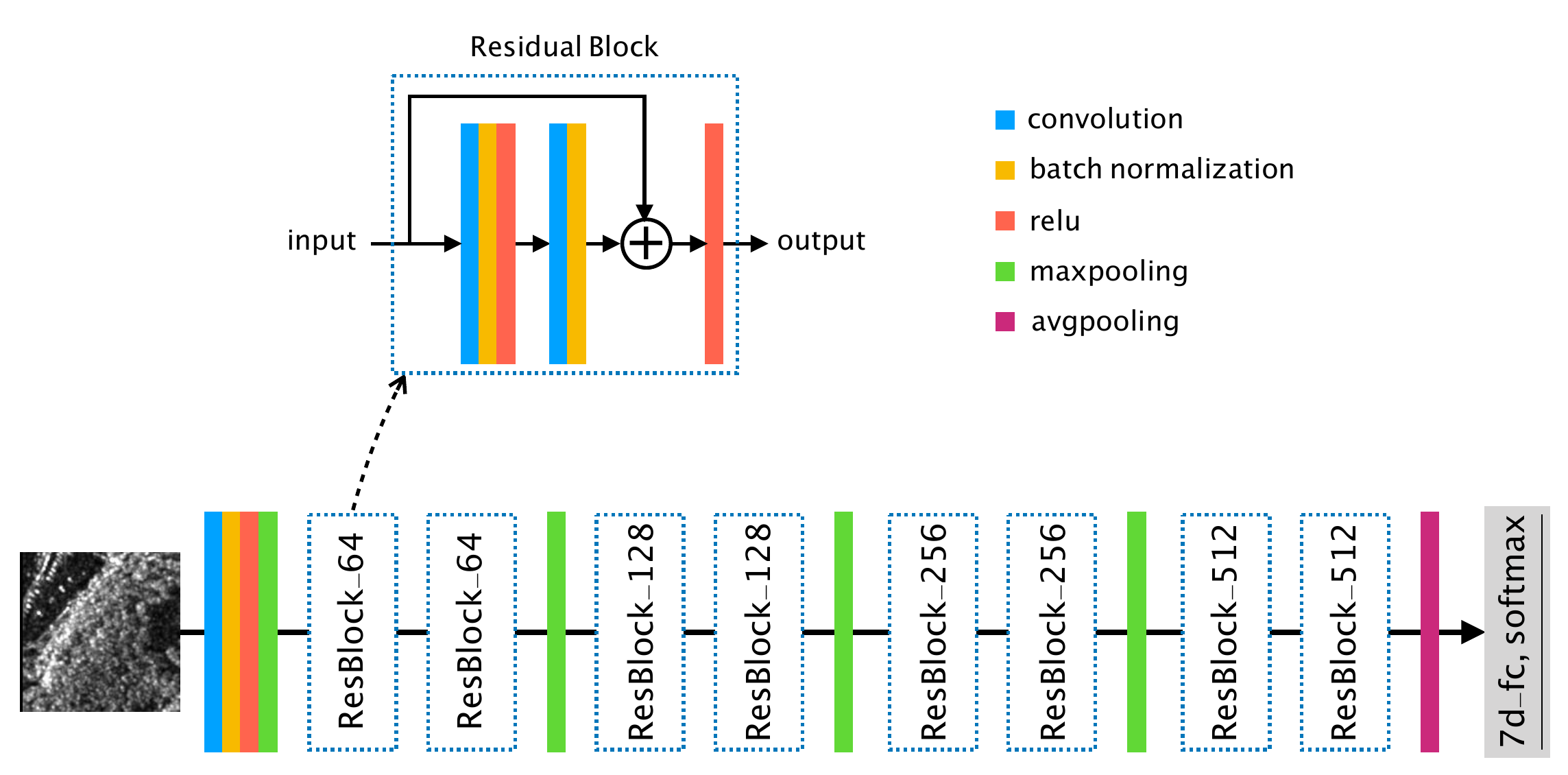}
\caption{Residual network architecture applied in our experiments.}
\label{network}
\end{figure}

\subsection{Transfer Learning with Deep Residual Networks}

ResNet-18 \cite{He} shown in Fig. \ref{network}, containing four residual blocks repeated twice, with two convolution layers in each block, followed by an average pooling layer and a fully-connected layer related to the class number of the task, was applied in our experiments. Such a very deep well-trained neural network (with a data volume of about 43.7 MB) has a great ability in extracting hierarchical features but is difficult to train from scratch. Transfer learning provides a better initialization to the parameters in deep layers, making the optimization easier and speeding up the training process. The ImageNet pre-trained model is popular as a transferred source. However, the large difference lying between SAR land cover patches and natural object images may harm the transferability, making it necessary to find a bridge to narrow the gap \cite{hzl2}. 

Rather than using the ImageNet pre-trained model as most remote sensing applications do, we proposed a transitive transfer learning method with the help of the NWPU-RESISC45 dataset \cite{NWPU} to build a bridge linking the natural and SAR land cover images. ResNet-18 pre-trained on ImageNet \cite{He} was applied as our initial model and then the convolution layers were fine-tuned with the NWPU-RESISC45 dataset, obtaining the remote sensing enhanced model as the transfer source for our applications.

\subsection{Imbalanced Classification}

When we trained a model with a dataset of highly imbalanced classes, we applied some techniques such as sampling and cost-sensitive learning \cite{kukar1998cost}. For evaluation, we considered some metrics beyond accuracy, like precision, recall and F1-score.

Each time when we input a batch of samples to the network, over-sampling of the minority classes and under-sampling of the majority classes were applied to ensure a balanced distribution in that batch. Moreover, the training samples were weighted according to the corresponding proportions of the dataset in order to penalize the misclassification of the minority classes more heavily than for the majority classes. 

\subsection{Top-2 Smooth Loss Function}
Here, we introduced the top-2 smooth loss function \cite{berrada2018smooth} to reduce the label noise in our dataset. With the classification of seven types of land cover in level-1, we mainly focused on the top-2 predictions $\mathrm{\mathbf{\bar{y}}} \in \mathbb{R}^2$. We define the label space $\mathcal{Y}=\{1,2,...,n\}$ and the ground truth label $y \in \mathcal{Y}$, the top-2 predictions $\mathrm{\mathbf{\bar{y}}} \in \mathcal{Y}^{(2)}$. The subset of $\mathcal{Y}^{(2)}$ where the ground truth $y$ is included, is written as $\mathcal{Y}_y^{(2)}$. The top-2 smooth loss function with temperature can be defined as: 

\begin{equation}
\begin{split}
L_{2, \tau}(\mathbf{s},y) = &\tau \log \lbrack \sum \limits_{\mathrm{\mathbf{\bar{y}}} \in \mathcal{Y}^{(2)}} \exp ( \frac{1}{\tau} (\Delta(\mathrm{\mathbf{\bar{y}}}, y) + \frac{1}{2} \sum \limits_{j \in \mathrm{\mathbf{\bar{y}}}} s_j )) \rbrack \\
& - \tau \log \lbrack \sum \limits_{\mathrm{\mathbf{\bar{y}}} \in \mathcal{Y}_y^{(2)}} \exp ( \frac{1}{2\tau} \sum \limits_{j \in \mathrm{\mathbf{\bar{y}}}} s_j ) \rbrack.
\end{split}
\end{equation}

where $\mathbf{s} \in \mathbb{R}^n$ denotes a vector of scores of $n$ labels, $\tau$ the temperature parameter and $\Delta(\mathrm{\mathbf{\bar{y}}}, y) \triangleq \mathbbm{1} (y \notin \mathbf{\bar{y}})$ denotes the margin. The top-2 smooth loss function takes two most likely predictions into consideration, making it possible to prevent over-fitting with label noise when learning only with the cross-entropy loss of a single label.

In the proposed method, the top-2 smooth loss was combined with the cross-entropy loss with a trade-off $\lambda$ rather than totally replacing it as \cite{berrada2018smooth} did, because $k=2$ is not much smaller than $n=7$ in our case. The top-2 loss was used as a constraint to alleviate model bias towards the single misleading label when there was more than one kind of land cover on the ground. Finally, adding an L2 regularization term and the loss weight $w_y$, the combined loss function can be written as: 

\begin{equation}
\begin{split}
L_{2, \tau, \mu, \lambda}(\mathbf{s}, y) = & (1 - \lambda) (-\sum \limits_i s_i \log \frac{\exp(s_i)}{\sum \limits_{j=1}^C \exp(s_j)} ) \\
& + w_y \lambda L_{2, \tau}(\mathbf{s},y) + \mu \sum \limits_i \| \beta_i \|^2,
\end{split}
\end{equation}
where $\beta_i$ denotes the weights in each layer, and $\lambda$, $\tau$ and $\mu$ are three hyper-parameters which control the trade-off between the top-2 and cross-entropy losses, temperature tuning, and the penalty of the regularization term. The cost weight $w_y$ related to the class $y$ is given by:

\begin{equation}
w_y = \frac{1}{n-1}(1 - \frac{N_y}{\sum_{i=1}^{n}{N_i}}),
\end{equation}
where $n$ denotes the class number and $N_i$ denotes the number of samples in the $i^{th}$ class.

In addition, we use bagging \cite{Bauer1999}, training five sub-models with re-sampling data and voting the predictions of all sub-models in order to make the trained model more robust and general.

\section{Experimental Results and Discussions}
\label{4}
We firstly introduce our experimental settings in Section \ref{4.1}. In Section \ref{4.2} and \ref{4.3} we analyze the effectiveness of different transfer sources and the combined top-2 smooth loss. Finally, we use three examples to demonstrate the good generalization of the proposed SAR specific deep model in \ref{4.4}.

\subsection{Experimental Settings}
\label{4.1}
On the one hand, we randomly chose 200 patches in each category from collection 01 to 36 as \textit{Test Data 1} and the remaining as \textit{Training Data}. On the other hand, patches of the full scene images from another seven collections were used as \textit{Test Data 2} which had a different distribution with \textit{Training Data} and \textit{Test Data 1}. The details of training and test data settings are presented in Table \ref{data}.

\begin{table*}[!t]
\caption{Training and Test Dataset}
\label{data}
\centering
\begin{tabular}{ccccccccc}
\toprule
\textbf{Category} & \textbf{Settlements} & \textbf{\tabincell{c}{Industrial\\areas}} & \textbf{\tabincell{c}{Public\\transportations}} & \textbf{\tabincell{c}{Agricultural\\land}} & \textbf{\tabincell{c}{Natural\\vegetation}} & \textbf{\tabincell{c}{Bare\\ground}} & \textbf{\tabincell{c}{Water\\bodies}} & \textbf{Total} \\
\midrule
\textit{Training Data} & 24,930 & 2,979 & 4,485 & 6,029 & 4,911 & 2,240 & 6,826 & \textbf{52,400} \\
\textit{Test Data 1} & 200 & 200 & 200 & 200 & 200 & 200 & 200 & \textbf{1,400} \\
\textit{Test Data 2} & 3,155 & 563 & 928 & 925 & 1,383 & 1,319 & 2,133 & \textbf{10,406} \\
\bottomrule
\end{tabular}
\end{table*}

The TSX detected images are represented by 16 bits spanning a large dynamic range. In our experiments, the 16 bit data after log transform was fed into the network to prevent the gradients from vanishing during training. When transferring the pre-trained model, only the last fully-connected layer was adapted to fit the new task.

The hyper-parameters of $\lambda$, $\tau$ and $\mu$ which control the trade-off between the top-2 and cross-entropy losses, the temperature tuning, and the penalty of the regularization term, were empirically set to 0.2, 1.0 and 0.25, respectively. Adam optimization algorithm \cite{adam} was applied and the initial learning rate was set to $10^{-5}$ when using the top-2 loss, and to $10^{-4}$ otherwise.

\subsection{The Transitive Transferring Model}
\label{4.2}
We compared the results of overall classification accuracy in \textit{Test Data 1} and the average stopping time of training a ResNet-18 network for the following cases to demonstrate the effectiveness of transferring from proper source data to SAR land cover classification:

\subsubsection{RD-$\mathbf{TSX}$}By randomly initializing all layers of ResNet-18, the network was trained with the TSX dataset from scratch.
\subsubsection{TL-$\mathbf{ImageNet}$2$\mathbf{TSX}$} By using the pre-trained model of ImageNet, the network was fine-tuned with the TSX dataset with only the last layer related to the classes trained from scratch.
\subsubsection{TL-$\mathbf{ImageNet}$2$\mathbf{RS}$2$\mathbf{TSX}$} The hierarchical convolution layers of ImageNet pre-trained model were fine-tuned with NWPU-RESISC45 optical remote sensing images for 45 types of land cover classification and then transferred to the TSX dataset.

\begin{figure}[!t]
\centering
\includegraphics[width=7cm]{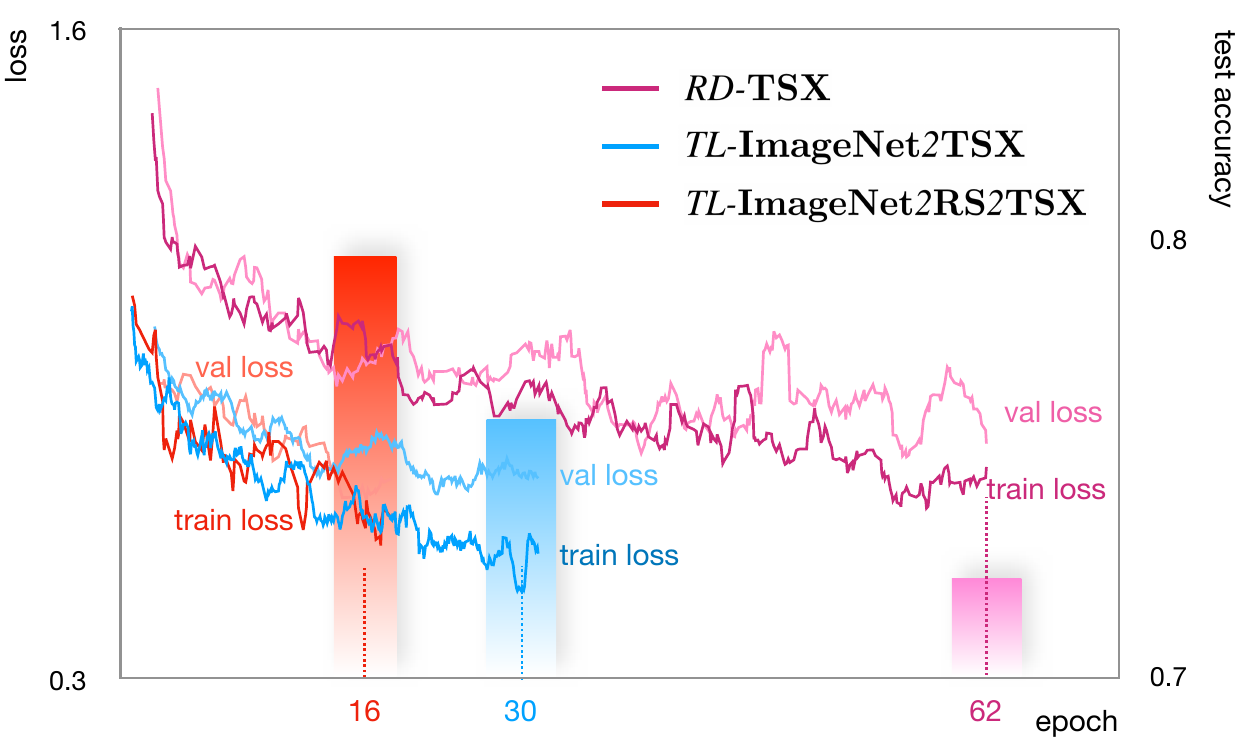}
\caption{The lines denote the loss values on training and validation data before early stopping, and the bars denote the validation accuracy of three experimental settings.}
\label{Fig5}
\end{figure}

Fig. \ref{Fig5} illustrates that training such a very deep network like ResNet-18 with the TSX dataset from scratch is difficult and time-consuming. Consequently, it is necessary to find a better initialization of millions of parameters. Although the ImageNet pre-trained model has been widely used for remote sensing applications, it is controversial to directly transfer the natural images to SAR data due to the huge differences between them. Here, we took advantage of the remote sensing images with similar categories to our SAR land cover dataset to narrow the gap between natural images and SAR data. We see a performance improvement both in test accuracy and training time. 

\subsection{Combined with Top-2 Loss}
\label{4.3}
Besides applying some strategies such as Dropout and L2 regularization to avoid over-fitting during training, we found that label noise may play an important role with a negative influence on the trained model. To address this prediction bias issue, we introduced a top-2 smooth loss function to constrain the cross-entropy from introducing a bias to the noisy labels. We use different metrics of top-1, top-2 overall accuracy, and also F1-score to evaluate the results which is defined by $F_1 = 2/(p^{-1} + r^{-1})$, where $p = \frac{TP}{TP+FP}$, $r = \frac{TP}{TP+FN}$, and $TP$, $FP$, $FN$ denote the true positive, false positive and false negative samples, respectively.

\renewcommand{\multirowsetup}{\centering} 
\begin{table}[!t]
\caption{Classification Result on Test Data 1 and Test Data 2}
\label{result}
\centering
\begin{tabular}{ccccc}
\toprule
  & \diagbox{\textbf{Eval}}{\textbf{Loss}} & \textbf{ce loss} & \textbf{top-2 loss} & \textbf{combined loss} \\
  \midrule
  \multirow{3}{*}{\textbf{\textit{Test Data 1}}}
        & \textbf{F1 score}  & 0.8318       & 0.7954  & \textbf{0.8327} \\
        & \textbf{top-1 acc} & 83.14\%      & 79.5\%  & \textbf{83.21\%} \\
        & \textbf{top-2 acc} & \textbf{95.36\%} & 93.29\%   & 95.00\% \\
  \midrule
  \multirow{3}{*}{\textbf{\textit{Test Data 2}}}
        & \textbf{F1 score}  & 0.709      & 0.684   & \textbf{0.72} \\
        & \textbf{top-1 acc} & 74.09\%      & 73.24\% & \textbf{76.78\%} \\
        & \textbf{top-2 acc} & 91.26\%      & 91.25\%   & \textbf{92.62\%} \\
  \midrule

  \multicolumn{2}{c}{\textbf{training time}} & 16 epochs & \textbf{6 epochs} & 9 epochs \\
\bottomrule
\end{tabular}
\end{table}

The top-2 smooth loss focuses on two most likely land cover classes, thus penalizing the predictions only when neither of the top-2 likely labels are the ground truth. Table \ref{result} shows the training time of using different loss functions and the test results of F1-score, top-1, and top-2 accuracies both on \textit{Test Data 1} and \textit{2}. The top-2 accuracy is the percentage of time that the classifier gives the correct class in the top-2 probabilities. We found that training with top-2 smooth loss only would lead to a fast convergence of the deep network to speed up training. However, the top-2 smooth loss was too relaxing for the optimization resulting in much smaller gradients after a period of time and weights being difficult to update. As a consequence, we combined the top-2 smooth loss with the cross-entropy as a constraint to achieve a faster convergence with nine epochs compared with 16 epochs of learning with cross-entropy. With the improvement in F1-score and accuracy, together with the obvious speed-up in training time, the combined loss performed considerably better.


\begin{figure}
\centering
\includegraphics[width=6cm]{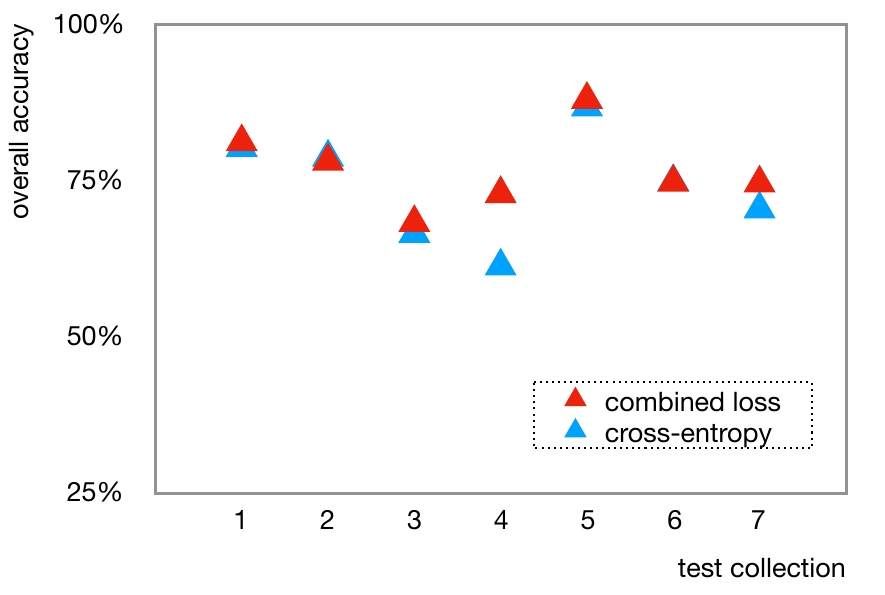}
\caption{The classification results on seven new collections of \textit{Test Data 2}, learning with cross-entropy, and the combined loss shown as blue/red dots, respectively.}
\label{Fig7}
\end{figure}

\subsection{Generalization to Other Data and Tasks}
\label{4.4}
We could demonstrate the good generalization of our SAR deep ResNet model by testing two tasks: (1) land cover classification on new collections of \textit{Test Data 2}; (2) MSTAR \cite{mstar} target recognition without data augmentation, as shown in Table \ref{result22}.


\begin{table}[!t]
\caption{Transfer to New Tasks on SAR Data}
\label{result22}
\centering
\begin{tabular}{cccc}
\toprule
\textbf{Dataset} & \textbf{Task} & \textbf{Classes} &\textbf{Transfer Result} \\
\midrule
\textit{Test Data 2} & scene classification & 7 & 76.78\%  \\
\textit{MSTAR \cite{mstar}} & target recognition & 10 & 99.46\%  \\
\bottomrule
\end{tabular}
\end{table}

Fig. \ref{Fig7} shows the test results of each new collection in \textit{Test Data 2} which has a different distribution with the training data. In most new collections, the performance of the combined loss was better than cross-entropy (2.69\% higher on average) which demonstrates that the trained model with combined loss reduces over-fitting due to label noise and improves generalization on new dataset. 

Furthermore, when we transferred the proposed model to an MSTAR \cite{mstar} 10-class recognition task, we achieved a state-of-the-art overall accuracy of 99.46\%, compared with 98.02\% using an ImageNet pre-trained model, and 99.09\% by other CNN-based methods \cite{chen2016target}.


\section{Conclusion}
\label{5}
In this letter, a deep transfer learning method combining the top-2 smooth loss was proposed to solve the land cover classification problem for a large-scale HR TSX dataset, with highly imbalanced classes, geographic diversity and noisy labels. Rather than applying the ImageNet pre-trained model of ResNet-18 to SAR images directly, we used an optical remote sensing land cover dataset to narrow the gap between SAR and natural images which results in a significant improvement on feature transferability. The proposed combined loss function is successful in reducing model bias to noisy labels and our proposed deep ResNet model shows a good generalization which allows it to be exploited for other SAR-specific tasks.



%



\section*{Acknowledgment}
We thank the TerraSAR-X Science Service System for the provision of images (Proposals MTH-1118 and LAN-3156), and University of Chinese Academy of Sciences (UCAS) Joint PhD Training Program for the provision of a scholarship. We would also thank Dr. Gottfried Schwarz for improving this letter.

\ifCLASSOPTIONcaptionsoff
  \newpage
\fi



\bibliographystyle{IEEEtran}
\bibliography{IEEEabrv,GRSL}
%



%







\end{document}